\documentclass[bm]{epl}

\newcommand{\kagome}{{kagom\'e} }

\usepackage{amsmath}
\usepackage{amssymb}

\title{Double-peak specific heat feature in frustrated antiferromagnetic clusters}
\author{A. V. Syromyatnikov \and S. V. Maleyev}
\institute{Petersburg Nuclear Physics Institute, Gatchina, St. Petersburg 188300, Russia}


\begin{document}

\maketitle

\begin{abstract}
We study the nature of the double-peak specific heat structure in \kagome clusters. That containing 12 spins is considered thoroughly by numerical diagonalization. Simple models are proposed revealing the low-$T$ peak nature at $T_l<\Delta$ ($\Delta$ is the spin gap) in this case and in those of larger clusters studied so far. We show that the rapid increase in density of states just above the spin gap gives rise to this peak. These models establish the reason for the weak magnetic field sensitivity of the low-$T$ peak. Our discussion could be appropriate for other frustrated antiferromagnetic systems too. 
\end{abstract}

{\it Introduction}.---Much attention in last decade was given to such frustrated magnets as \kagome and pyrochlore. This interest was motivated by their unusual low-$T$ magnetic properties observed experimentally (see~\cite{canals,ramirez,hiroi} and references therein). For instance, the deviation of their uniform susceptibility from Curie-Weis law occurs at much lower temperatures than for non-frustrated ones. Then the specific heat $C$ measurements in spin-$\frac32$ \kagome material SrCrGaO revealed a peak at $T\approx 5$ K which was practically independent of magnetic field up to 12 T and $C\propto T^2$ at $T\lesssim 5$ K \cite{ramirez}. For qualitative understanding of these systems low-$T$ physics numerous finite clusters diagonalization studies were carried out \cite{elser,elstner,sind,lecheminant,zeng,leung,wald,zeng2,canals,kawa}.

In particular, investigations of \kagome Heisenberg spin-$\frac12$ clusters with the number of sites $N\le36$ and with periodic boundary conditions \cite{lecheminant,zeng,leung,wald,zeng2} revealed a gap $\Delta$ separating the ground state from the upper triplet levels and a band of nonmagnetic singlet excitations inside the spin gap. It was obtained that the number of states in the singlet band increases exponentially with the number of sites. The similar picture with many singlets inside the spin gap exists in pyrochlore clusters too \cite{canals}.

Specific heat calculations of \kagome clusters with $N=12$, 18, 24 and 36 revealed two peaks with the low-$T$ one at $T_l\lesssim\Delta$ \cite{zeng2,elser,elstner,sind}. It was obtained for $N=18$, 36 in~\cite{sind} that the low-$T$ peak is weakly dependent on magnetic field and $C\propto T^2$ in a small interval below it. The point of view is accepted now that the wealth of low-lying singlet excitations is responsible for the low-$T$ peak \cite{elstner,sind} and for its field independence \cite{ramirez,sind}. At the same time it was pointed out in~\cite{sind,elser} that upper triplet levels contribute to it as well.

The double-peak specific heat structure is a general feature of many frustrated antiferromagnetic systems. It was obtained numerically in Heisenberg spin-$\frac12$ pyrochlore slab \cite{kawa}, in $\Delta$ chain \cite{Japanese} and in clusters of triangular lattice in the model with multi-spins exchange taken into account \cite{misg,roger}. The last approach was proposed to describe the second layer of $^3$He absorbed on graphite \cite{roger}.

\begin{figure}
		\onefigure{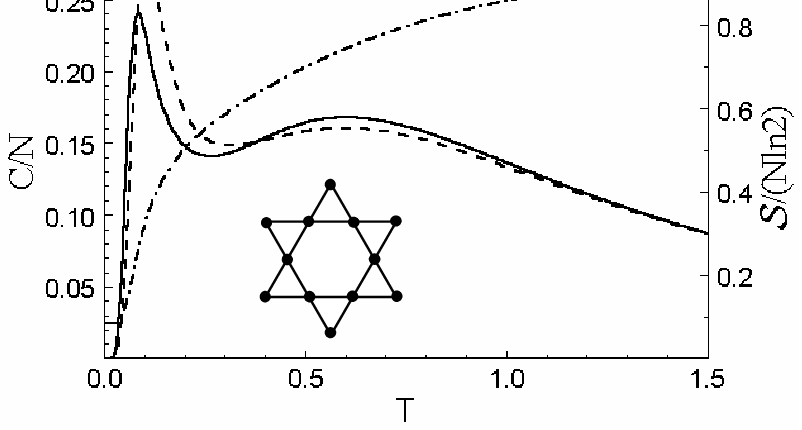}
		 \caption
{
Specific heat $C(T)$ per spin (solid line) and entropy ${\cal S}(T)$ divided by ${\cal S}(\infty)=N\ln2$ (dashed dotted line) of the antiferromagnetic cluster with $N=12$ shown below the curves (star). The specific heat calculated with eq.~(\ref{c}) is shown by dashed line.
}
\label{heat}
\end{figure}

In the present paper we study the nature of the double-peak specific heat structure in Heisenberg spin-$\frac12$ \kagome antiferromagnetic clusters. Our conclusions on the origin of the low-$T$ peak contradict to those having been proposed in the previous works \cite{elstner,sind}. We start with numerical diagonalization of the \kagome Heisenberg spin-$\frac12$ cluster plotted in the inset of Fig.~\ref{heat} (star). As is shown in our recent papers~\cite{syromyat} the star has doubly degenerate singlet ground state separated from the lower triplet level by the gap $\Delta$. The lower singlet band obtained in the previous finite clusters studies \cite{lecheminant,zeng,leung,wald,zeng2} appears in this approach from the stars degenerate ground states as a result of inter-star interaction.

So there are no singlet states inside the spin gap in the star as it was in clusters with periodic boundary conditions studied so far. Nevertheless we have obtained double-peak structure of $C$ with the low-$T$ peak at $T_l<\Delta$ which possesses weak field sensitivity similar to results of the previous studies \cite{zeng2,elser,elstner,sind}. Simple models revealing the low-$T$ peak nature is proposed for the \kagome star and other larger antiferromagnetic clusters. It is shown that the rapid increase in density of states just above the gap is responsible for this peak. These models establish the reason for the weak magnetic field sensitivity of the low-$T$ peak.

The nature of the double-peak structure in other frustrated antiferromagnetic systems can be the same as it is for \kagome clusters considered here. We hope that the present paper will stimulate the corresponding studies.

{\it Analysis of the star}.---We begin with consideration of the Heisenberg spin-$\frac12$ antiferromagnetic cluster shown in the inset of Fig.~\ref{heat} which Hamiltonian has the form
\begin{equation}
\label{h}
{\cal H} = \sum_{\langle i,j \rangle}{\bf S}_i{\bf S}_j-H\sum_iS_i^z,
\end{equation}
where $\langle i,j\rangle$ denote nearest neighbors and the value of exchange $J=1$ so as temperature $T$ and magnetic field $H$ to be measured in coupling constant unite. We start with $H=0$ in eq.~(\ref{h}). As the Hamiltonian commutes with all projections of the total spin operator, all star levels are classified by the values $S$, irreducible representations (IRs) of its symmetry group $C_{6v}$ and are degenerated with $S^z$. In the basis of IRs the matrix of the Hamiltonian has a block structure. Each block has been diagonalized numerically. Some low-lying levels are presented in Table~\ref{levels}.

\begin{table}
\caption{
Low-lying levels of the star. They are classified by $S$. There are no levels with $S>2$ in the energy sector presented.
}
\label{levels}
\begin{center}
\begin{largetabular}{cccc}
	\multicolumn{1}{c}{Energies}&\multicolumn{3}{c}{Number of levels}\\
	& $S=2$ & $S=1$ & $S=0$ \\
	\hline
	-4.500000 & 0 & 0 & 2 \\
	-4.240331 & 0 & 1 & 0 \\
	-4.236220 & 0 & 2 & 0 \\
	-4.232400 & 0 & 2 & 0 \\
	-4.202448 & 0 & 1 & 0 \\
	-4.183814 & 0 & 0 & 1 \\
	-4.182320 & 0 & 0 & 2 \\
	-4.141850 & 0 & 0 & 2 \\
	-4.077928 & 0 & 1 & 0 \\
	-4.068850 & 0 & 2 & 0 \\
	-4.056472 & 0 & 0 & 1 \\
	-4.010310 & 0 & 2 & 0 \\
	-3.913465 & 0 & 1 & 0 \\
	-3.865010 & 0 & 0 & 2 \\
	-3.832691 & 0 & 1 & 0 \\
	-3.829460 & 1 & 0 & 0 \\
\end{largetabular}
\end{center}
\end{table}

As is seen the star has doubly degenerate singlet ground state separated from the lower triplet level by the spin gap $\Delta\approx0.26$ \cite{syromyat}. Energies of the ground state and the upper level are $-4.5$ and 4.5, respectively. A distinguishing characteristic of the spectrum in the range $\Delta-4.5<E<4.5$ is that levels are very close to each other: distances between them are of the order of $0.1\div0.01\ll\Delta$. A part of the spectrum shown in Table~\ref{levels} reflects this feature. We show below that this peculiarity plays the crucial role for low-$T$ star properties.

Specific heat $C$ calculated with this spectrum is shown in Fig.~\ref{heat}. It has a double-peak structure with low-$T$ peak at $T_l\approx0.085$ and high-$T$ one at $T_h\approx0.6$. The most intriguing feature of $C$ is the existence of the low-$T$ peak at $T_l<\Delta$. The matter is that according to the common point of view specific heat of a fully gaped system should decrease exponentially with $T$ in the range $T<\Delta$. Meanwhile in our case the exponential decay occurs at $T<T_l$ only. Moreover in a usual situation a peak in $C$ results from a peak in the DS. As is demonstrated below, it is not the case for the star. We show now that the extremely high DS just above the gap is in the origin of this surprising behavior.

\begin{figure}
		\onefigure{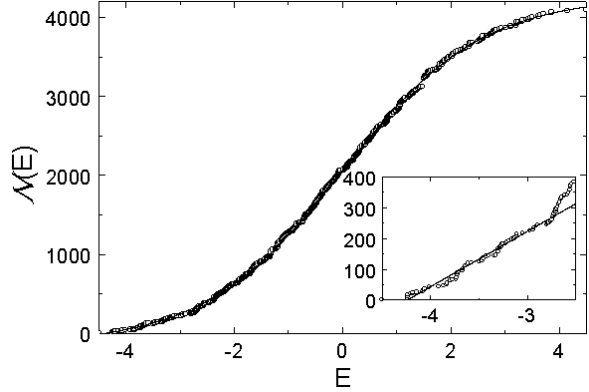}
		 \caption
{
\label{density_fig}
Integrated density of states of the star ${\cal N}(E)$ which is the number of states with energies lower than $E$. It is fitted by $2200\tanh(0.39E)+2060$. The inset shows the low-energy sector where ${\cal N}(E)$ is fitted by $180(E+4.24)$.
}
\end{figure}

A graphic of integrated density of states (IDS) ${\cal N}(E)$ which is the number of states with energies lower than $E$ is presented in Fig.~\ref{density_fig}. It is well fitted by the function ${\cal N}(E)=2200\tanh(0.39E)+2060$. As distances between levels at $\Delta-4.5<E<4.5$ are of the order of $0.1\div0.01$ one can use this function as IDS above the gap for the specific heat modeling in the range $T\gg0.01$. Then the DS given by $\rho(E)=d{\cal N}(E)/dE$ can be approximated on the interval $\Delta-4.5<E<4.5$ as follows:
\begin{equation}
\rho(E) = \frac{w}{\cosh^2(gE)},
\label{density}
\end{equation}
where $w=858$ and $g=0.39$. In this model $C$ has the form
\begin{equation}
\label{c}
C=\frac{d\overline{E}}{dT}=\frac{d}{dT}\left(\frac{\int_\Delta^9 dE e^{-E/T}E\rho(E-4.5)}{d+\int_\Delta^9 dE e^{-E/T}\rho(E-4.5)}\right),
\end{equation}
where $\rho(E)$ is given by eq.~(\ref{density}) and $d$ is degeneracy of the ground state which is equal to 2 for the star. In eq.~(\ref{c}) and below we shift the energy scale for convenience so as the ground state energy to be equal to 0. Specific heat calculated from eqs.~(\ref{density}) and (\ref{c}) by numerical integration is also presented in Fig.~\ref{heat}. It has a double-peak structure and coincides well with that obtained with the real spectrum. It reproduces the high-$T$ peak which evidently corresponds to the maximum of the DS at $E=0$ and has the low-$T$ peak at $T_l=0.1$ which is 15\% higher than the real one. In any way the result reproduces the characteristic features of the specific heat and encouraged us to use this model in further $C$ analysis.

{\it Low-$T$ peak}.---Let us consider in detail the specific heat at $T\lesssim\Delta$. We derive now an analytic expression for it in this case in the assumption $\frac1\Delta\gg g$ which is held for the star ($\frac1\Delta\approx4$ and $g\approx0.4$). It is easy to show using eq.~(\ref{density}) that at these conditions one can replace $\rho(E-4.5)$ in eq.~(\ref{c}) by a constant $W\approx\rho(\Delta -4.5)\approx117$. Unfortunately in this low-energy region approximation eq.~(\ref{density}) is not very good. As is shown in Fig.~\ref{density_fig}, an accurate value of $W$ is 180 because the real ${\cal N}(E)$ is well fitted by $180(E+4.24)$ in the range $\Delta-4.5<E<-2.7$. As a result we have for the specific heat at $T\lesssim\Delta$ from eq.~(\ref{c}):
\begin{equation}
\label{heateq}
C=\frac{1+Zxe^x(x^2+2x+2)}{(1+Zxe^x)^2},
\end{equation}
where $x=\Delta/T$ and $Z=d/(W\Delta)$. Position of the peak $T_l$ is determined by the condition $dC/dx=0$ which has the form
\begin{equation}
\label{equation}
Ze^x=\frac{x+4}{x^2+2}.
\end{equation}
As follows from eq.~(\ref{equation}) the peak appears if $Z<Z_c\approx2$. For the star $Z\approx0.04$ and numerical solution of eq.~(\ref{equation}) gives $T_l=0.093$ which is very close to the real value of 0.085. Analysis of eqs.~(\ref{heateq}) and (\ref{equation}) shows that the peak becomes smaller and $T_l$ tends to zero as $\Delta$ decreases. At the same time $T_l$ becomes larger as $W$ decreases and even exceeds the value of $\Delta$ when $W$ is small enough. In any way the peak disappears as soon as $Z$ amounts to $Z_c$. So the value of $\Delta$ and that of the jump in the DS above the gap are responsible for the peculiar low-$T$ star properties.

\begin{figure}
\onefigure{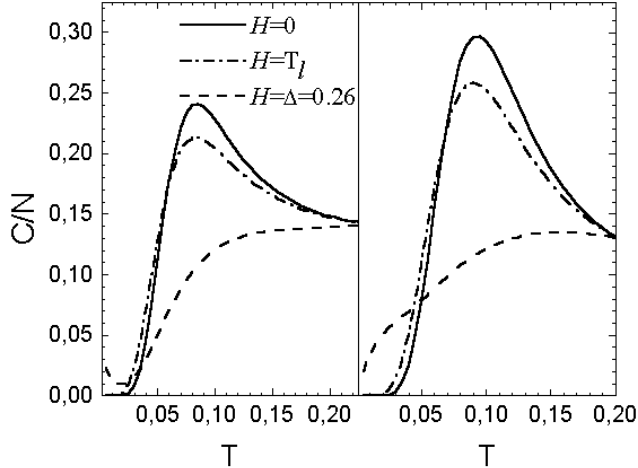}
\caption
{
\label{speak}
Low-$T$ peak evolution for the star in the magnetic field $H$ obtained by numerical diagonalization of the Hamiltonian eq.~(\ref{h}) (left) and by the model discussed in the text (right). In both cases $T_l\approx\Delta/3$.
}
\end{figure}

{\it Field dependence}.---We turn now to the discussion of the low-$T$ peak dependence on the external magnetic field $H$. In an isotropic unfrustrated antiferromagnet a singularity in $C/T$ implies transition to long-range order. It is suppressed to zero temperature by $H$ of the value of the peak temperature. In this respect the field behavior of
the star at small $T$ is also unusual. The results of $C$ calculations using the spectrum of the Hamiltonian eq.~(\ref{h}) obtained numerically are presented in the left graph of Fig.~\ref{speak}. It is seen that at $H=T_l\approx0.085$ the peak height is reduced 11\% only and $T_l$ is diminished slightly. The field of the value of the spin gap is needed to wash out the peak.

The low-$T$ peak evolution can be described by the model discussed above. It allows also to understand the nature of this weak $H$ sensitivity. The field splits levels with $S\ne0$. It is clear from Table~\ref{levels} that the low-energy sector above the gap is represented mostly by triplets. So the DS can be modeled at $H<\Delta$ in the following way: it is 0 at $0<E<\Delta-H$, the DS is $W/3$ at $\Delta-H<E<\Delta$, it is $2W/3$ at $\Delta<E<\Delta+H$ and the DS remains $W$ at  $E>\Delta+H$. The results of the peak evolution with $H$ in this approach are presented in the right graph of Fig.~\ref{speak}. It is seen that this model reproduces even quantitatively the main tendencies in low-$T$ peak evolution with $H$. The field smooths the area where the drop in DS takes place and in doing so it causes the peak reduction. {\it For the changes in the peak to be significant $H\approx\Delta$ is needed. So the reason for the weak field sensitivity of the low-$T$ peak in the star is that $T_l\approx\Delta/3$.}

{\it Larger \kagome clusters}.---As was mentioned above the double-peak specific heat structure in \kagome clusters with even $N$ was obtained in many previous numerical works \cite{zeng,zeng2,elser,elstner,sind}. Height and position with $T_h\approx2/3$ of the high-$T$ peak got by high-temperature expansion \cite{elstner} and by finite cluster diagonalization studies \cite{zeng,zeng2,elser,elstner,sind} coincide with each other and with our results. The low-$T$ peak with $T_l\lesssim\Delta$ was obtained in~\cite{zeng2,elser,elstner,sind} but its position and height were depended on $N$, cluster form and calculation technique. We show now that the reason for the low-$T$ peak appearance in those cases is similar to that for the star discussed above: rapid increase in the DS. It will be argued that the low field sensitivity of the low-$T$ peak obtained in~\cite{sind} has the same origin as it is for the star.

We focus here on the largest cluster with $N=36$ studied before which low-energy spectrum was obtained in~\cite{wald,zeng2} and thermodynamical properties were discussed in~\cite{sind}. According to~\cite{wald} the spectrum above $\Delta$ is nearly continuous as it is for the star. But in contrast the spin gap there is filled with singlets separated from the non-degenerate singlet ground state by a very small gap $\Delta_s\ll\Delta$. According to results of~\cite{wald} distribution of these singlets is quite even with DS of the order of 1000. The DS above the gap is much larger than that inside it.

Qualitative understanding of the low-$T$ peak nature in this case can be obtained by the model similar to that proposed above for the star. Let us assume that the DS is $W'$ at $E<\Delta'$, where $\Delta'\gtrsim\Delta$, and there is a jump in the DS at $E=\Delta'$ so as it is equal to $W\gg W'$ above $\Delta'$. Specific heat calculation based on the formula similar to eq.~(\ref{c}) gives on the interval of interest $\Delta_s\ll T\lesssim\Delta'$ the following equation:
\begin{equation}
\label{heatsin}
C=\frac{(1+re^x)^2+rx^2e^x+2rZxe^{2x}+Zxe^x(x^2+2x+2)}{(1+re^x+Zxe^x)^2},
\end{equation}
where now $x=\Delta'/T$, $Z=1/(\Delta' W)$ and $r=W'/W$. We have made an effort to reproduce the low-$T$ peak in $C/N$ obtained in~\cite{sind} with the height of approximately 0.152 and $T_l\approx0.05<\Delta\approx0.074$. eq.~(\ref{heatsin}) at $\Delta'=0.23$, $W=43000$ and $W'=1000$ gives a peak with these parameters but the approach is too rough to get $T^2$ behavior of $C$ at $T<T_l$. As is demonstrated below there are no more sets of parameters in this model describing the peak discussed in~\cite{sind}. In this case $Z\Delta'/T_l\ll r$ and terms in eq.~(\ref{heatsin}) containing $Z$ do not affect the low-$T$ peak. As a result $T_l$ is given by the equation
\begin{equation}
\label{eqsin}
re^x=\frac{1}{x-1}.
\end{equation}
So the low-$T$ peak properties are determined by $\Delta'$ and the ratio $r$. Thus they are a result of cooperative effect of states above and below $\Delta'$. This finding is in accordance with that of~\cite{sind}.

It is seen from eqs.~(\ref{heatsin}) and (\ref{eqsin}) that value of $\Delta'$ does not affect the peak height but it determines $T_l$. Evidently $T_l$ decreases as $\Delta'$ tends to zero. The value of $r$ determines both the peak position and its height. The peak becomes smaller and $T_l$ increases as $r$ becomes larger. So it is clear now that if, according to~\cite{wald}, DS below $\Delta'$ is chosen equal to 1000 and $W'\ll W$ the values of $\Delta'$ and $W$ are determined unambiguously by the low-$T$ peak obtained in~\cite{sind}.

Using this model it is easy to describe qualitatively the weak sensitivity of $C$ to the magnetic field which was discussed in~\cite{sind}. It was demonstrated there that at $H=T_l=0.05$ the peak is decreased by 10\% only and $T_l$ is also decreased slightly. The situation here is similar to that for the star considered above. As magnetic field splits levels with $S\ne0$ the DS in the field $H<\Delta'$ will be the following: the DS at $0<E<\Delta'-H$ remains $W'$, it is $W'+W/3$ at $\Delta'-H<E<\Delta'$, the DS is $2W/3$ at $\Delta'<E<\Delta'+H$ and the DS remains $W$ above $\Delta'+H$. We also assume here that the spectrum above $\Delta'$ at $H=0$ is formed mostly by triplets. As a result of $C$ calculations in our model it was obtained that the field $H=T_l=0.05$ reduces the low-$T$ peak by 8\% and diminishes $T_l$ slightly. These findings are in even quantitative egreement with those of~\cite{sind}. The reduction of the peak at $H=\Delta'/2$ calculated in our model is about 28\%. So we see that field $H\sim\Delta'\gg T_l$ is needed to change the low-$T$ peak significantly.

{\it Entropy analysis}.---Entropy of the star $\cal S$ is also presented in Fig.~\ref{heat}. Its temperature dependence obeys the most characteristic feature of those of the larger clusters: in all cases there is 50\% of the total entropy in the low-$T$ peak at $T<0.2$ \cite{sind,elstner}. It is seen from the above consideration that this pecularity in \kagome clusters stems from the rapid increas in DS and states above $\Delta'$ give the main contribution to ${\cal S}$ at $T<0.2$. Such an entropy feature was obtained in other models of frustrated antiferromagnets mentioned above and having double-peak $C$ structure \cite{misg,roger}.

In closing we point out that the low-$T$ peak in $C$ can remain in spin-$\frac12$ \kagome antiferromagnets. The peak at small $T$ with weak field sensitivity obtained experimentally in SrCrGaO \cite{ramirez} could be of the same nature.

{\it Conclusion}.---In this paper we study the nature of the double-peak specific heat structure in \kagome clusters. That containing 12 spins and shown in the inset of Fig.~\ref{heat} (star) is considered thoroughly by numerical diagonalization. Simple models are proposed revealing the low-$T$ peak nature at $T_l<\Delta$ ($\Delta$ is the spin gap) in this case and in those of larger clusters studied so far numerically \cite{zeng2,elser,elstner,sind}. We show that the rapid increase in density of states just above the gap gives rise to this peak. These models establish the reason for the weak magnetic field sensitivity of the low-$T$ peak. Our consideration could be appropriate for other frustrated antiferromagnetic systems too and, as we hope, will stimulate further studies.

\acknowledgments

This work was supported by the Russian State Program "Collective and Quantum Effects in Condensed Matter", the Russian Foundation for Basic Research (Grant Nos. 03-02-17340, 00-15-96814) and the Russian State Program "Quantum Macrophysics".

\end{document}